\documentclass[letter]{aa} 

\usepackage{graphicx}
\usepackage{txfonts}

\usepackage{float}
\usepackage{hyperref}
\hypersetup{breaklinks=true,colorlinks=true,urlcolor=blue, linkcolor=blue,  citecolor=blue, bookmarksopen=true}
\usepackage{placeins}
\usepackage{tikz}
\usepackage{xcolor}
\usepackage{siunitx}

\begin{document}

\title{T CrA has a companion\thanks{Based on observations collected at the European Southern Observatory under ESO programmes 111.253T.001, 113.26PG.001, 113.26PG.006, and 113.26PG.008.}}
\subtitle{First direct detection of T CrA B with VLTI/MATISSE}

   \author{
	J.~Varga\inst{\ref{inst_K},\ref{inst_MT}} \and
	A.~Matter\inst{\ref{inst_O}} \and
	F.~Millour\inst{\ref{inst_O}} \and
	G.~Weigelt\inst{\ref{inst_B}} \and
	R.~van~Boekel\inst{\ref{inst_H}} \and
	B.~Lopez\inst{\ref{inst_O}} \and
	F.~Lykou\inst{\ref{inst_K},\ref{inst_MT}} \and
	Á~Kóspál\inst{\ref{inst_K},\ref{inst_MT},\ref{inst_EL},\ref{inst_H}} \and
	L.~Chen\inst{\ref{inst_K},\ref{inst_MT}} \and
	P.~A.~Boley\inst{\ref{inst_H}} \and
	S.~Wolf\inst{\ref{inst_Ki}} \and
	M.~Hogerheijde\inst{\ref{inst_L},\ref{inst_P}} \and
	A.~Moór\inst{\ref{inst_K},\ref{inst_MT}} \and
	P.~Ábrahám\inst{\ref{inst_K},\ref{inst_MT},\ref{inst_EL},\ref{inst_Vi}} \and
	J.-C.~Augereau\inst{\ref{inst_I}} \and
	F.~Cruz-Saenz de Miera\inst{\ref{inst_To},\ref{inst_K},\ref{inst_MT}} \and
	W.-C.~Danchi\inst{\ref{inst_Go}} \and
	Th.~Henning\inst{\ref{inst_H}} \and
	T.~Juhász\inst{\ref{inst_K},\ref{inst_EL},\ref{inst_MT}} \and
	P.~Priolet\inst{\ref{inst_I}} \and
	M.~Scheuck\inst{\ref{inst_H}} \and
	J.~Scigliuto\inst{\ref{inst_O}} \and
	L.~van~Haastere\inst{\ref{inst_L}} \and
	L.~Zwicky\inst{\ref{inst_K},\ref{inst_MT}}
      	}  
   \institute{
Konkoly Observatory, Research Centre for Astronomy and Earth Sciences, HUN-REN, Konkoly-Thege Mikl\'os \'ut 15-17, 1121 Budapest, Hungary\label{inst_K}\\ \email{varga.jozsef@csfk.org} \and                                           	 
	CSFK, MTA Centre of Excellence, Konkoly-Thege Miklós út 15-17, H-1121 Budapest, Hungary\label{inst_MT} \and
	Université Côte d’Azur, Observatoire de la Côte d’Azur, CNRS, Laboratoire Lagrange, France\label{inst_O} \and
	Max-Planck-Institut für Radioastronomie, Auf dem Hügel 69, 53121, Bonn, Germany\label{inst_B} \and
	Max Planck Institute for Astronomy, K\"onigstuhl 17, D-69117 Heidelberg, Germany\label{inst_H} \and
	ELTE E\"otv\"os Lor\'and University, Institute of Physics, P\'azm\'any P\'eter s\'et\'any 1/A, 1117 Budapest, Hungary\label{inst_EL} \and
	Institute of Theoretical Physics and Astrophysics, University of Kiel, Leibnizstr. 15, 24118 Kiel, Germany\label{inst_Ki} \and
	Leiden Observatory, Leiden University, PO Box 9513, 2300 RA Leiden, the Netherlands\label{inst_L} \and
	Anton Pannekoek Institute for Astronomy, University of Amsterdam, the Netherlands\label{inst_P} \and
	Institute for Astronomy (IfA), University of Vienna, T\"urkenschanzstrasse 17, A-1180 Vienna, Austria\label{inst_Vi} \and
	Univ. Grenoble Alpes, CNRS, IPAG, 38000 Grenoble, France\label{inst_I} \and
	Institut de Recherche en Astrophysique et Planétologie, Université de Toulouse, UT3-PS, CNRS, CNES, 9 av. du Colonel Roche, 31028 Toulouse Cedex 4, France\label{inst_To} \and
	NASA Goddard Space Flight Center, Astrophysics Division, Greenbelt, MD 20771, USA\label{inst_Go} 
 	}

   \date{Received September 15, 1996; accepted March 16, 1997}

 
  \abstract
{T CrA is a Herbig Ae-type young star in a complex circumstellar environment; it includes a circumstellar disk, accretion streamers, jets, and outflows. It has long been suspected to be a binary. However, until now, there has been no direct detection of a companion. Here we present new VLTI/MATISSE $L$- and $N$-band observations of T CrA taken between 2023 May and 2024 August with the aim of testing the binary nature of the system. We modeled the data with a geometric model using the Python tool \texttt{oimodeler}.
We detected a companion (T CrA B) with a projected separation of $\Delta r = 153.2 \pm 1.2$~mas ($\approx$$23$~au) toward the  west direction at a position angle of $275.4 \pm 0.1\degr$, in 2024 May--August.
Our results support that the companion has a nearly edge-on orbit that is highly misaligned with respect to the circumprimary disk. Such a configuration could cause warping and tearing of the disk around the primary, which has been proposed by recent studies. In the $L$ band the companion is extended, with a full width at half maximum (FWHM) size of $\sim$$1$~au, suggesting that the emission comes from a disk around the secondary star. The companion flux is $0.2$--$0.3$~Jy in the $L$ band, and $0.2$--$0.7$~Jy in the $N$ band, accounting for $4$--$20\%$ of the total emission at those wavelengths. The SED of the companion is compatible with thermal radiation of warm dust ($600$--$800$~K).
}


   \maketitle
%

\section{Introduction}

T CrA is an F0 spectral type young star in the Coronet Cluster of the Corona Australis star-forming region, at a distance of
$149.4\pm0.4$~pc \citep{Galli2020}.\footnote{The average distance of the Corona Australis region.} The star is encircled by a protoplanetary disk that was first detected at millimeter wavelengths by \cite{Henning1994}, and since then it has been imaged by ALMA, NACO, and SPHERE. The ALMA continuum image at $1.3$~mm at an angular resolution of $\approx$$0\farcs32$ shows a marginally resolved object with a full width at half maximum (FWHM) size of $0\farcs57 \pm 0\farcs03$ toward the major axis, and $0\farcs37 \pm 0\farcs02$ toward the minor axis, indicating an inclined disk \citep{Cazzoletti2019}. The position angle (PA) of the major axis was found to be $20.3 \pm 4.3\degr$ (measured east of north).
\cite{Cugno2023} detected the disk around T CrA with NACO in the $L'$ band ($3.5$--\SI{4.1}{\um}). They derived an inclination of $77 \pm 2\degr$ and a PA of $5 \pm 2\degr$. The disk appears as a ring with a radius of $0\farcs27$ ($40$~au, rescaled to the distance estimate of \citealp{Galli2020}).
\cite{Rigliaco2023} presented a SPHERE/IRDIS polarized  light image in the $H$ band, showing an extremely complex circumstellar environment. The outer disk appears to be almost edge-on and is seen as a dark lane. On the east side of the dark lane, inside a radius of $60$~au ($\sim$$0\farcs4$), a bright region is visible that can be interpreted as the disk surface facing the observer. The authors estimated a disk PA of $7 \pm 2\degr$ and an inclination of $85$--$90\degr$. 

T CrA has long been supposed to be a multiple system; however, until now, there has been no direct detection of any companion. From an analysis of spectro-astrometric data, \cite{Bailey1998} concluded that T CrA should be a binary, with a separation $>$$0\farcs14$ and a PA of $275\degr$. They added that the companion should be much fainter than the primary in the near-infrared (NIR), since IR speckle observations by \cite{Ghez1997} and \cite{Leinert1997} were unable to detect it.
In another spectro-astrometric study, \cite{Takami2003} also suggested that T CrA is a binary. They found that the PA of the spectro-astrometric displacement increases by $\sim$$2\degr/\mathrm{yr}$, consistent with the orbital motion of a companion. Based on this, they estimated a binary separation of $\sim$$0\farcs26$. 
They noted that the non-detection by \cite{Ghez1997} implies that the companion is fainter than $K=10.5$~mag. In a recent spectro-astrometric analysis, \cite{Whelan2023} also found evidence for the multiplicity of the T CrA system. They stated that the origin of east-west jets in their data is a putative companion lying at a separation of at least $30$~au ($0\farcs2$) and at a PA of $274\degr$. 
\cite{Rigliaco2023} found a sinusoidal variability pattern in the optical light curve of T CrA, with a period of $29.6$~yr and a $V$-band amplitude of $\sim$$1.4$~mag, which they attribute to the binary nature of the system. Based on a binary model, they suggest a mass of $1.7\ M_\sun$ for the primary and $0.9\ M_\sun$ for the secondary. 

In this letter we present new $L$- and $N$-band VLTI/MATISSE observations on T CrA, which we use to unambiguously confirm the binary nature of the source.


\section{Observations and data processing}
\label{sec:obs}

We observed T CrA with the MATISSE instrument on the Very Large Telescope Interferometer (VLTI) at the ESO Paranal Observatory in Chile, in the frame of our Guaranteed Time Observations (GTO) program. By combining the light of four telescopes, MATISSE delivers spectrally resolved interferometric data in the $L$  ($3$--\SI{4}{\um}), $M$  ($4.6$--\SI{5}{\um}), and $N$  ($8$--\SI{13}{\um}) bands \citep{Lopez2022}.
MATISSE samples the visibilities and closure phases of the object's brightness distribution on the sky.
The typical angular resolution is $\sim$$3$~mas in the $L$ band and $\sim$$10$~mas in the $N$ band. The field of view (FOV) of MATISSE is set by a pinhole that lets through only the central part of the beam into the instrument. The diameter of the pinhole is $1.5 \lambda/D$ in the $L$ band (at \SI{3.5}{\um}), and $2 \lambda/D$ in the $N$ band (at \SI{10.5}{\um}), where $D$ is the telescope diameter. This translates to a FOV of $130$~mas on the $8.2$~m diameter unit telescopes (UTs) and $600$~mas on the $1.8$~m diameter auxiliary telescopes (ATs) in the $L$ band, and $500$~mas on the UTs and $2\farcs3$ on the ATs in the $N$ band.
A typical MATISSE observation sequence starts with four non-chopped exposures, followed by eight chopped exposures. In the $LM$ bands the interferometric observables (correlated flux,\footnote{The correlated flux is the visibility in flux units, i.e., not divided by the single-dish flux.} closure phase), and single-dish flux data are obtained at the same time, while in the $N$ band the interferometric data are recorded in the non-chopped exposures, and the single-dish flux data in the chopped exposures.

Six observations of T CrA were obtained between 2023 May and 2024 August with MATISSE, using various instrument modes and telescope configurations. An overview is shown in Table \ref{tab:obs}. All  of the data sets except one (2024 June) were taken with the GRA4MAT fringe tracker mode \citep{Woillez2024}. The chopped exposure sequence was used in all observations except one (2023 May). We processed the data with the standard MATISSE pipeline DRS ver. 2.0.2.
Post-processing Python tools were also used for flux calibration\footnote{\url{https://github.com/Matisse-Consortium/tools/}} and to average exposures. For our analysis, we used the $L$- and $N$-band non-chopped correlated flux (when available), non-chopped closure phase, and chopped single-dish flux. The $M$-band data were discarded from the modeling because of the lower signal-to-noise ratio of the correlated flux ($\sim$$50$) and the relatively large noise on the closure phase ($\sim$$2\degr$). The 2024 Aug 10 observation is incomplete, and was taken without a calibrator. However, for that observation we can still use the $L$-band closure phase.
In the $N$ band, T CrA is not bright enough for the ATs, and thus only the UT data is usable. We were unable to obtain $N$-band single-dish spectra in either of the UT epochs, and thus we used instead an archival Spitzer spectrum taken on 2008 May 06 (AORkey: 21884416, \citealp{Lebouteiller2011}\footnote{The Spitzer spectrum was obtained from the LR7 version of the CASSIS spectral atlas.}). 
Our MATISSE data have a spectral resolution $\lambda/\Delta\lambda\sim$$30$ (LOW), except the 2023 May $L$-band data, which have $\lambda/\Delta\lambda\sim$$500$ (MEDIUM).

\section{A first look at the data}
\label{sec:lookdata}

Our final data products are shown in Figs.~\ref{fig:data_L}, \ref{fig:data_L_uncal}, and \ref{fig:data_N}. The most striking aspect is the clear presence of sinusoidal modulations in the $L$-band AT, and $N$-band UT interferometric data, indicative of multiple distinct objects in the FOV of the instrument. It is also notable that the $L$-band UT data do not show such modulations. We argue that it is because there is just a single source within the UT $L$-band FOV, while there are multiple centers of emission in the AT $L$-band and UT $N$-band FOV.
This implies a separation of more than $65$~mas and less than $250$~mas; this assumption is tested later in our modeling.
Overall, the correlated flux decreases with increasing spatial frequency (i.e., baseline length), indicating that the brightness distribution is not point-like, but is somewhat extended. With a simple Gaussian fit to the UT data, we get a FWHM size of $\sim$$2$~mas for the $L$-band emitting region of the central source.  The $N$-band Spitzer spectrum shows a silicate spectral feature in emission. The silicate feature can be also seen in some of the $N$-band MATISSE correlated spectra, although it is less clear because of the strong binary modulation signal.

 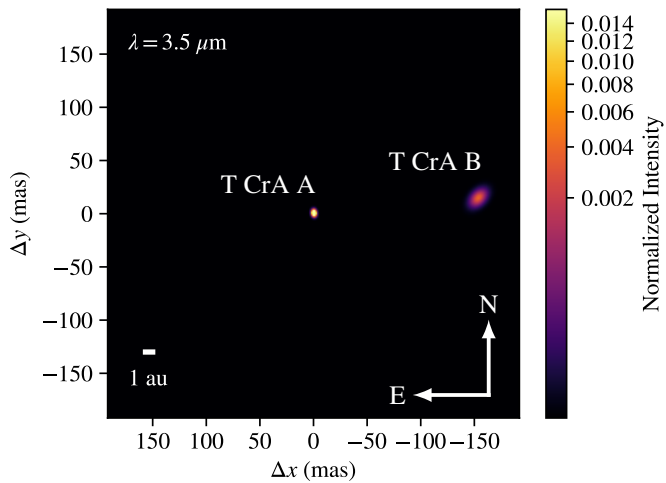
\begin{figure}
	\center
	\begin{tikzpicture}[line width=2pt,line cap=rect]
	\node (image) at (0,0) {
\includegraphics[width=\hsize,trim={0cm 0.3cm 0cm 0.23cm},clip]{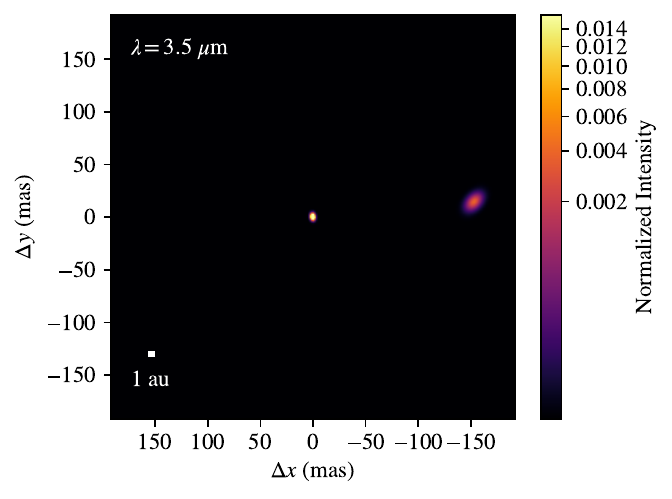}
};
    	\node at (-0.9,0.8) {\color{white} T CrA A};
    	\node at (1.3,1.1) {\color{white} T CrA B};
    	\draw [line width=0.5mm,white,-latex](2,-2) -- (1,-2) ;
    	\draw [line width=0.5mm,white,-latex](2,-2) -- (2,-1) ;
    	\node at (0.8,-2) {\color{white} E};
    	\node at (2,-0.8) {\color{white} N};
\end{tikzpicture}
\caption{Best-fit model image corresponding to the fit to the three complete $L$-band data sets. The companion is clearly visible at $153$~mas separation toward the west direction. The J2000 coordinate of the central source is $\alpha = 19^\mathrm{h}01^\mathrm{m}58.79^\mathrm{s}$, $\delta = -36\degr57\arcmin50.34\arcsec$.}
     	\label{fig:model_img}
\end{figure}

\section{Modeling}
\label{sec:model}

To extract spatial information from the data, we performed modeling with simple geometric shapes, using the \texttt{oimodeler}\footnote{\url{ https://github.com/oimodeler/oimodeler}} Python package \citep{Meilland2024}. To find the best fit, we employed Markov chain Monte Carlo (MCMC) sampling with \texttt{emcee} \citep{Foreman-Mackey2013b}. Our model consists of a central point source (the primary star), two elongated 2D Gaussians, and a uniform background component. One of the Gaussians (the inner disk of the primary) is centered on the point source, while the other (the companion) is allowed to be at an offset position within the instrument FOV. We use the term companion to denote the system composed of the secondary star and its circumstellar disk. Each Gaussian has three parameters: the FWHM (in the major axis direction), the minor/major axis ratio ($b/a$), and the PA of the major axis ($\theta$). Assuming that the structures we see are inclined disks, $b/a$ is linked to the disk inclination ($i$) by $\cos i = b/a$. The uniform background accounts for an extended emission component which is over-resolved even at the shortest baselines of our data. The model components are chromatic. For the point source, we used the model spectrum of the central star from the spectral energy distribution (SED) modeling of \cite{Rigliaco2023}. For the other components, the $L$-band flux contributions were fitted at the edges of the used wavelength range ($3.1$--\SI{3.9}{\um}), and the intermediate values were linearly interpolated. In the $N$ band, we modeled the flux contributions with a combination of a linear continuum and a silicate spectral template for each component individually. The template is extracted from the Spitzer spectrum of T CrA. Three fitted parameters correspond to this flux model, the continuum fluxes at the edges of the fitted wavelength range ($8$--\SI{13}{\um}) and the peak flux value in the silicate template. The FWHMs of the Gaussians are achromatic in the $L$ band, while they are wavelength-dependent in the $N$ band, using the same linear trend plus silicate template parameterization that we use for the flux contributions.

During the initial fits, we found that the modulation signal of the companion was often poorly reproduced. We attribute this partly to absolute calibration errors which can shift the spectra up or down \citep[for more details, we refer to Appendix B in][]{Varga2021}, and partly to our choice for the radial brightness profile of the model components (Gaussian), which may be too simplistic. The sinusoidal modulations in the interferometric spectra are directly related to the companion's position, while the overall level of the correlated spectra at different baselines is mainly determined by the components' surface brightness distributions. Thus, absolute errors will mainly affect the model parameters describing the radial profiles, and channel-to-channel noise will affect the companion coordinates. In order to be able to perform an accurate fit to the spectral modulation signal, we do the following. 	First, to account for the absolute errors, we introduce correction factors ($C_{\mathrm{corr}}$),  which are multiplied by the model correlated fluxes on a per-baseline basis or on a per-epoch basis. Each of these factors is a fitted parameter. Then we replace the pipeline produced error bars, which contain both absolute and random errors, with the random noise estimated from the correlated flux and closure phase data itself.

We found that the typical random noise is $0.01$~Jy (correlated flux) and $0.4\degr$ (closure phase) in the $L$-band data, and $0.015$~Jy (correlated flux) and $0.6$--$1\degr$ (closure phase) in the $N$-band data. For the $L$-band single-dish spectra, we used the original error bars provided by the pipeline, while in the $N$ band we calculated the error as the standard deviation of four spectra\footnote{The following spectra were used: An ISOPHOT-S spectrum from 1996 Apr 6 \citep{kospal_atlas}, a VLTI/MIDI spectrum from 2004 Aug 1 \citep{Varga2018}, and two Spitzer spectra from 2006 Oct 23 and 2008 May 6 (AORkeys 16828416 and 21884416).} taken at different epochs, thus representing the uncertainty, amounting to $\sim$$15\%$, due to the temporal variability of the object's emission. The uncertainty of the $N$-band single-dish spectrum only affects the parameters on the flux contributions, all other parameters, including the position of the companion, are not influenced by that error source.
A further source of uncertainty in the positional fit is the spectral calibration accuracy of MATISSE data, which is estimated to be $\sim$$0.5\%$\footnote{This estimate was provided to us by the MATISSE consortium (priv. communication), based on the results of the MATISSE commissioning (ESO document VLT-TRE-MAT-15860-9141 v1.0).} in LOW spectral resolution. This propagates to the companion coordinates, causing an uncertainty of the same level. We include this $0.5\%$ uncertainty in the error budget of our position measurements.

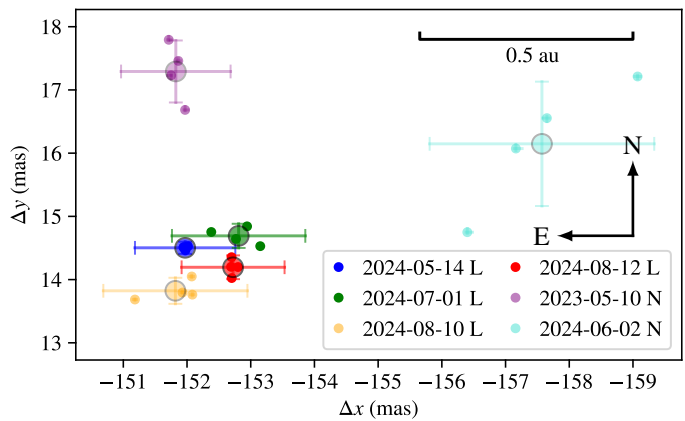
\begin{figure}
	\center
	\begin{tikzpicture}[line width=2pt,line cap=rect]
    	\node (image) at (0,0) {
	\includegraphics[width=\hsize,trim={0cm 0.2cm 0cm 0.05cm},clip]{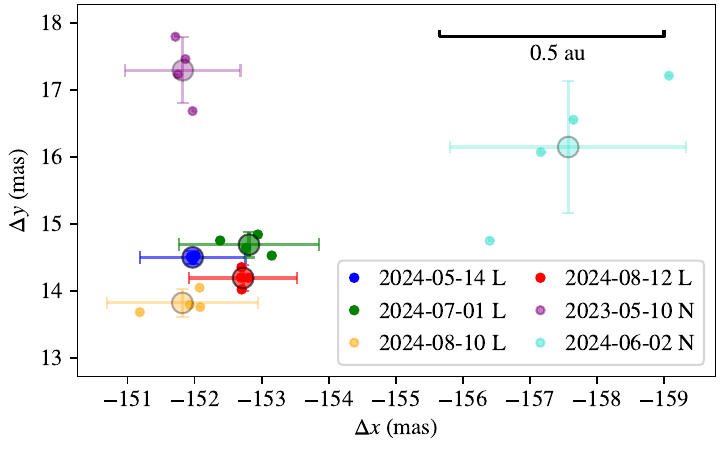}
	};
        	\draw [line width=0.35mm,-latex](3.8,-0.3) -- (2.8,-0.3) ;
        	\draw [line width=0.35mm,-latex](3.8,-0.3) -- (3.8,0.7) ;
        	\node at (2.6,-0.3) {E};
        	\node at (3.8,0.9) {N};
	\end{tikzpicture}

	\caption{Relative position of the companion, T CrA B, with respect to the central star, shown at the various epochs of our data. The small circles indicate the per-exposure fit results; the large circles show the per-epoch averages.  
	}
         	\label{fig:xy}
\end{figure}

\begin{table}
	\caption[]{\label{tab:fit_res}Overview of the fitted parameters from our modeling. The parameters on the position of the companion are reported separately in Table~\ref{tab:fit_xy}.}
	\centering
	\small
   	\begin{tabular}{lp{0.5cm}p{1.1cm}p{2.0cm}}
   \hline \hline
   & Unit & Stage 1 & Stage 2 \\
   \hline
   \multicolumn{4}{c}{\it Central source } \\
   \hline
	FWHM ($L$ band) & mas \newline au & $2.8$ \newline $0.4$ &
  	$2.1^{+0.5}_{-0.7}$  \newline $0.3^{+0.1}_{-0.1}$ \\[1mm]
	FWHM (\SI{11}{\um}) & mas \newline au &  &
 	$12.8^{+1.1}_{-0.9}$ \newline $1.9^{+0.2}_{-0.1}$\\[1mm]
   $\cos\,i$ & & $0.64$ \\
   $\theta$\tablefootmark{(a)} & $\degr$ & $6$ \\
   $F_\nu$ (\SI{3.5}{\um}) & Jy & $2.3$ & $1.8^{+0.3}_{-0.4}$ \\[1mm]
   $F_\nu$ (\SI{11}{\um}) &  Jy & & $3.1^{+0.4}_{-0.2}$ \\[1mm]
   \hline
   \multicolumn{4}{c}{\it Companion} \\
   \hline
   FWHM ($L$ band) & mas \newline au &
   $9.1$ \newline $1.4$ &
   $6.2^{+0.6}_{-0.8}$ \newline $0.9^{+0.1}_{-0.1}$ \\[1mm]
   FWHM (\SI{11}{\um}) & mas \newline au &  &
   $19.0^{+2.6}_{-1.8}$ \newline $2.8^{+0.4}_{-0.3}$\\[1mm]
   $\cos\,i$ & & $0.64$ \\
   $\theta$ & $\degr$ & $138$ \\
   $F_\nu$ (\SI{3.5}{\um}) &  Jy & $0.4$ & $0.22^{+0.08}_{-0.06}$ \\[1mm]
   $F_\nu$ (\SI{11}{\um}) & Jy & & $0.6^{+0.1}_{-0.1}$ \\[1mm]
   \hline
   \multicolumn{4}{c}{\it Background} \\
   \hline
   $F_\nu$ (\SI{3.5}{\um}) &  Jy & $0.001$ & $0.6^{+0.5}_{-0.5}$ \\[1mm]
   $F_\nu$ (\SI{11}{\um}) & Jy & & $1.8^{+0.4}_{-0.4}$ \\[1mm]
	\hline
	$\chi^2_\mathrm{red}$ ($L$ band) & & 714.5 & $16.2$ -- $41.8$\tablefootmark{(b)}\\[1mm]
	$\chi^2_\mathrm{red}$ ($N$ band) & & & $23.8$ -- $59.3$\tablefootmark{(b)}\\[1mm]
   \hline
   	\end{tabular}
   \tablefoot{
   \tablefoottext{a}{This parameter was kept fixed during the fitting.}
   \tablefoottext{b}{The range of $\chi^2_\mathrm{red}$ values corresponding to the separate per-exposure fits is indicated here.}
   }
\end{table}

We employed a two-stage fitting approach. In the first stage, we fitted the three complete $L$-band AT data sets at the same time to get a general view on the system configuration. Then, in the second stage, we performed separate fits to each exposure in order to obtain the position of the companion independently for every epoch. To this end, we averaged the best-fit positions of the four exposures in each epoch. The errors were estimated as the standard deviation of the four resulting positions combined with the wavelength calibration error. In the case of the FWHMs of the Gaussians and the flux contributions, we calculated their mean values averaged over multiple epochs,\footnote{For the averaging, we used the epochs 2024 May 14, 2024 Jul 1, and 2024 Aug 12 for the $L$ band, and the epochs 2023 May 10 and 2024 Jun 2 for the $N$ band.} and the error bars were determined by taking the $16$th and $84$th percentile ranges of the sample values. The fitting procedure is explained in more detail in the Appendix \ref{sec:app_model}.

\section{Results and discussion}
\label{sec:res}

We present our first-stage fits to the three complete $L$-band AT data sets in Fig.~\ref{fig:fit_3epochs}. The fits indicate that the closure phases and the single-dish flux are relatively well represented by the model, including the sinusoidal modulation signal. The fits to the correlated fluxes are less sufficient. The $L$-band model image of the best fit (Fig.~\ref{fig:model_img}) shows a clearly visible companion at $153$~mas separation toward the west ($\mathrm{PA} = 275.4\degr$). This corresponds to a projected separation of $22.9$~au at a distance of $149.4$~pc. The fitted parameters are listed in Table~\ref{tab:fit_res}.

The results from the second-stage fits are reported in Table~\ref{tab:fit_res} (FWHMs and flux contributions) and Table~\ref{tab:fit_xy} (companion positions). Furthermore,   
Table~\ref{tab:fit_LN_res} gives a more detailed overview on the wavelength dependence of the FWHMs and flux contributions in both bands. Data--model comparison plots are shown in Fig.~\ref{fig:fit_CF_CP}. The second-stage fits have much lower $\chi^2_\mathrm{red}$ values compared to the first stage, with typical values of $\sim$$25$ in the $L$ band and $\sim$$40$ in the $N$ band. Figure~\ref{fig:fit_CF_CP} shows that the frequency and phase of the binary modulation signal are adequately modeled, but there is some mismatch in the amplitude of the signal between the data and the model. The cause of the mismatch may be related to the spatial structure of the components, to chromatic effects, or both.
The mean of the $C_{\mathrm{corr}}$ values is $1.05$, and their standard deviation is $0.24$. Absolute errors, which are expected to be in the $5$--$20\%$ range \citep{Varga2021}, partially account for this variance. The fact that the model sometimes requires $C_{\mathrm{corr}}$ values larger than the typical absolute error also suggests that the  spatial structure of the components is more complex than that our model can represent.

\begin{table}
	\caption[]{\label{tab:fit_xy}Fit results on the position of the companion.}
	\centering
	\small
   	\begin{tabular}{lllll}
\hline \hline
Date & $\Delta x$ & $\Delta y$ & Unit & Band \\
\hline
\multicolumn{5}{c}{\it Fits to individual epochs}\\
\hline
2023-05-10 & $-151.82 \pm 0.86$ & $17.29 \pm 0.49$ & mas & $N$\\
2024-05-14 & $-151.97 \pm 0.79$ & $14.50 \pm 0.10$ & mas & $L$\\
2024-06-02 & $-157.57 \pm 1.76$ & $16.15 \pm 0.98$ & mas & $N$\\
2024-07-01 & $-152.81 \pm 1.05$ & $14.69 \pm 0.19$ & mas & $L$\\
2024-08-10 & $-151.82 \pm 1.13$ & $13.82 \pm 0.21$ & mas & $L$\\
2024-08-12 & $-152.72 \pm 0.81$ & $14.19 \pm 0.19$ & mas & $L$\\
\hline
\multicolumn{5}{c}{\it Grand average 2024 May -- 2024 Aug}\\
\hline
  & $-152.50 \pm 1.17$ & $14.46 \pm 0.30$ &  mas & $L$\\
  & $ -22.78 \pm 0.18$ & $2.16 \pm 0.04$  &   au & $L$\\
\hline
\end{tabular}
\end{table}

\begin{figure}
	\center
	\includegraphics[width=\hsize,trim={0cm 0.3cm 0cm 0.22cm},clip]{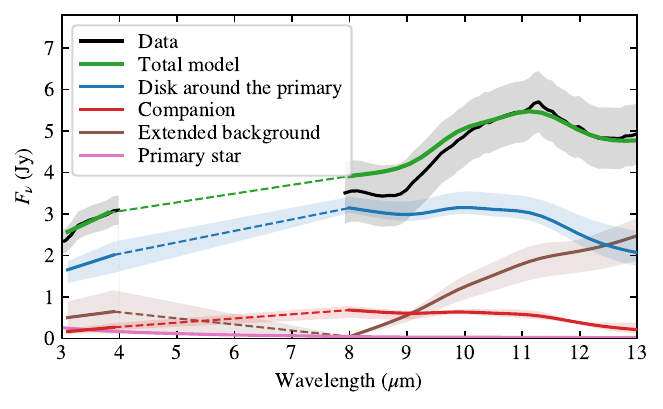}
	\caption{Our fit to the $L$- and $N$-band SED of T CrA. The $L$-band data are the average of the three MATISSE AT single-dish spectra taken in 2024, while the $N$-band spectrum was taken in 2008 by Spitzer. The SED of the central star is extracted from the SED modeling of \citet{Rigliaco2023}. The dashed lines indicate the model SED interpolated between the $L$ and $N$ bands.}
         	\label{fig:SED}
\end{figure}
    
The $L$-band model image (Fig.~\ref{fig:model_img}) shows a relatively compact object in the center, in agreement with our preliminary fits to the UT $L$-band data (Sect.~\ref{sec:lookdata}). In contrast, the companion is more extended, with a FWHM of $6$--$9$~mas ($\sim$$1$~au). This is a robust detection that we verified with a modified model in which the Gaussian of the companion is replaced by a point source. This model provides a much worse fit because the  high-frequency spectral modulations it produces have significantly larger amplitudes than those in the data. Thus, we can exclude that the companion is a compact unresolved object at thermal IR wavelengths. We propose that the extended emission originates from a moderately inclined disk around the secondary star which would   otherwise be too faint to be detected in our data. That is also supported by the $N$-band fits, which indicate a characteristic size of the emitting region of $15$--$20$~mas ($2$--$3$~au) in the center of the band.

The position of the companion is well constrained by the data, as shown in Fig.~\ref{fig:xy} and in Table~\ref{tab:fit_xy}. The $L$-band derived positions are within a $1$ mas radius circle around the average position of $\Delta x = -152.5$~mas, $\Delta y = 14.5$~mas. We checked the robustness of our positional fits against different choices for the radial surface brightness profile of the components. One model had pseudo-Lorentzian brightness profiles \citep[as in][]{Lazareff2017}, the other had uniform ellipsoidal disks. We found that the difference of the best-fit postions from these models compared to our baseline model with Gaussians are $<$$0.07$~mas in the $L$ band and $<$$0.35$~mas in the $N$ band, less than the error bars of the per-epoch fits.

The PA of the companion ($275.4 \pm 0.1\degr$) is in excellent agreement with the values inferred by the spectro-astrometric studies of \citet{Bailey1998} and \citet{Whelan2023}. The apparent invariability of the companion PA implies an almost edge-on orbit that is highly misaligned with respect to the circumprimary disk.
Curiously, the companion positions inferred from the $N$-band fits lie several mas away from the $L$-band points. Proper motion cannot explain these offsets, as the 2024 June position should lie between the 2024 May and 2024 July points, but this is not the case. A possible explanation is that the photocenter of the emission is shifted between the $L$- and $N$-band wavelengths, which might arise due to inclination effects in the circumprimary disk. To check whether the detected object is gravitationally bound, we looked up the proper motion data for T CrA. The typical values reported are $\sim$$20$--$30$~mas~yr$^{-1}$ \citep[e.g.,][]{Zacharias2013_UCAC4}, much larger than the largest possible relative proper motion of $6$~mas~yr$^{-1}$ based on our data. This hints at the two objects having a common proper motion. Another piece of evidence for a gravitationally bound system is the nearly constant PA of the companion over decades.

\citet{Rigliaco2023} presented two models featuring a binary on an edge-on circular orbit to explain the sinusoidal variability pattern in the optical light curve of T CrA. From their models, we calculated the expected separation for our MATISSE epochs. The model with an orbital period $P=29.6$~yr suggests a separation of $40$--$60$~mas in $2023$--$2024$, while the model with $P=59.2$~yr suggests $120$--$130$~mas. While neither of their models agrees with our position within our measurement errors, the model with the long period provides a relatively good match. Furthermore, in the model with the short period ($P=29.6$~yr), the maximum separation is only $\sim$$80$~mas, which is incompatible with our result of $153.2$~mas.

The flux contributions of the various components of the T~CrA system are relatively well constrained by our modeling, as shown in Tables \ref{tab:fit_res} and \ref{tab:fit_LN_res}, and Fig.~\ref{fig:SED}. The modeling suggests that the silicate spectral emission mainly comes from the circumprimary disk. The companion flux is in the range of $0.2$--$0.3$~Jy in the $L$ band, and $0.2$--$0.7$~Jy in the $N$ band. The flux ratio of the companion to the total emission is $7$--$9\%$ in the $L$ band and $4$--$20\%$ in the $N$ band. The $3$--\SI{13}{\um} SED of the companion suggests thermal radiation of warm dust ($\sim$$600$--$800$~K), likely located in the circumstellar disk of the secondary star.

The companion is at a position outside of the coronagraphic mask in the $H$-band SPHERE/IRDIS image of \citet{Rigliaco2023}. Interestingly, the location is a bit darker than the neighboring regions in their image, and is without any conspicuous features. On the other hand, the same location in the NACO $L'$-band image by \citet{Cugno2023} is occupied by a large ($\sim$$20$~au diameter) and relatively bright patch of emission. The non-detection of the companion in the near-IR may be due to dust obscuration, while the same material might be responsible for the extended $L'$-band emission seen in the NACO image, and for the overresolved emission component in our MATISSE data. It is unclear what the physical connection is between that extended structure and the companion detected in our data. A possible scenario was presented by \citet{Rigliaco2023}, in that the gravitational pull of the secondary star on a misaligned orbit creates an intermediate circumbinary disk, which may correspond to the east-west extended structure in the NACO and SPHERE images.

\section{Conclusion and summary}
\label{sec:concl}

We presented the first detection of the warm dust emission around the companion of T CrA, using IR interferometric observations with the VLTI/MATISSE instrument. The relative position of the companion, $\Delta r = 153.2$~mas, $\mathrm{PA} = 275.4\degr$, is very well constrained by our $L$-band data, with an uncertainty of $\sim$$1$~mas.
The position of the companion indicates that its orbit is misaligned with respect to the plane of the circumprimary disk. Such a misalignment can have dramatic effects on the disk, inducing warps and disk tearing, and is likely an important factor in shaping the complex circumstellar environment of T CrA. More VLTI observations are needed to track the orbital motion of the companion. A good orbital solution would be important for the understanding of how the companion interacts with the circumprimary disk. On a last note, deeper observations with high angular resolution direct imaging facilities might also be able to detect the companion and would help  shed more light on the dynamical processes in the circumstellar environment of T CrA. 

\begin{acknowledgements}
	MATISSE was designed, funded, and built in close collaboration with ESO, by a consortium composed of institutes in France (J.-L. Lagrange Laboratory -- INSU-CNRS -- C\^ote d’Azur Observatory -- University of C\^ote d'Azur), Germany (MPIA, MPIfR, and University of Kiel), the Netherlands (NOVA and University of Leiden), and Austria (University of Vienna). The Konkoly Observatory and Cologne University have also provided some support in the manufacture of the instrument.

	JV, P\'A, and FL are funded from the Hungarian NKFIH OTKA projects no. K-132406, and K-147380. 
	This work was also supported by the NKFIH NKKP grant ADVANCED 149943 and the NKFIH excellence grant TKP2021-NKTA-64. Project no.149943 has been implemented with the support provided by the Ministry of Culture and Innovation of Hungary from the National Research, Development and Innovation Fund, financed under the NKKP ADVANCED funding scheme.
	JV acknowledges support from the Fizeau exchange visitors programme. The research leading to these results has received funding from the European Union’s Horizon 2020 research and innovation programme under Grant Agreement 101004719 (ORP).
	FCSM received financial support from the European Research Council (ERC) under the European Union’s Horizon 2020 research and innovation programme (ERC Starting Grant “Chemtrip”, grant agreement No 949278). 

	The open-source \texttt{oimodeler} package used in this research is developed with support from the VLTI/MATISSE consortium and the ANR project MASSIF, and we would like to thank the whole development team. Extra thanks to Anthony Meilland, Marten Scheuck and Alexis Matter for their work.

	This research has made use of the services of the ESO Science Archive Facility.
    
	The Combined Atlas of Sources with Spitzer IRS Spectra (CASSIS) is a product of the IRS instrument team, supported by NASA and JPL. CASSIS is supported by the "Programme National de Physique Stellaire" (PNPS) of CNRS/INSU co-funded by CEA and CNES and through the "Programme National Physique et Chimie du Milieu Interstellaire" (PCMI) of CNRS/INSU with INC/INP co-funded by CEA and CNES.

\end{acknowledgements}

%
%

\bibliographystyle{aa}
\bibliography{ref_YSO_JV}

\begin{appendix}
   	 
\section{MATISSE observations and data}
\label{sec:app_dataplots}

\begin{table}[hbt!]
	\begin{minipage}{0.99\textwidth}    
	\caption{Overview of the MATISSE observations of T CrA. $\tau_0 $ is the atmospheric coherence time. LDD is the estimated angular diameter of the calibrator from \citet{Bourges2014}.}
	\begin{center}
    	\label{tab:obs}

	\begin{tabular}{l c c c c c c c c}
        	\hline
        	\hline
        	\multicolumn{5}{c}{Target} & \multicolumn{3}{c}{Calibrator} &  Grade\tablefootmark{a} \\
        	\hline
        	Date and time & Seeing & $\tau_0$ & Stations & Array & Name & LDD & Time & \\ 
        	(UTC) & ($''$) & (ms) &  &  & & (mas)& (UTC) & \\
        	\hline
	2023-05-10T10:14\tablefootmark{b} & 0.5 & 7.2 & U1-U2-U3-U4 & UTs & 29 Sgr & 2.2 & 09:54 & -- \\ 
	2024-05-14T07:29\tablefootmark{c} & 0.7 & 4.0 & A0-G2-J2-J3 & medium -- large & HD 184996 & 2.7 & 06:52 & C \\ 
	2024-06-02T04:30\tablefootmark{d} & 0.6 & 6.4 & U1-U2-U3-U4 & UTs & HD 181925 & 2.3 & 04:03 & A\\
	2024-07-01T02:24 & 0.4 & 6.7 & A0-G1-J2-K0 & large  & bet CrA & 2.5 & 02:50 & A\\
	2024-08-10T04:24\tablefootmark{e} & 1.0 & 4.2 & K0-G1-D0-J3 & large -- medium & -- & -- & -- & -- \\
	2024-08-12T03:43\tablefootmark{f} & 1.2 & 5.4 & A0-B5-D0-J3 & medium -- extended & bet CrA & 2.5 & 04:16 & C \\
    
        	\hline
    	\end{tabular}
	\end{center}
	\tablefoot{
	\tablefoottext{a}{The grade, given by the night astronomer in service mode (SM) runs, is a data quality indicator. 'A' is the best grade, 'B' means acceptable, and 'C' indicates that the data were recorded in conditions outside of the requested observing constraints.}
	\tablefoottext{b}{No grade, as this observation was in visitor mode. The data quality is good. Only non-chopped data were recorded, and thus this data set does not contain $N$-band single-dish flux data.}
	\tablefoottext{c}{Calibrator is at significantly different airmass.}
	\tablefoottext{d}{$N$-band single-dish spectrum could not be processed because of frame flagging issues during chopping.}
	\tablefoottext{e}{Observation was aborted because of a technical issue.}
	\tablefoottext{f}{Seeing was outside of the requested constraints, hence the grade C.}
	}
\end{minipage}  
	\end{table}

\begin{figure}[hbt!]
	\centering
	\begin{minipage}{0.96\textwidth}  
\includegraphics[width=\hsize]{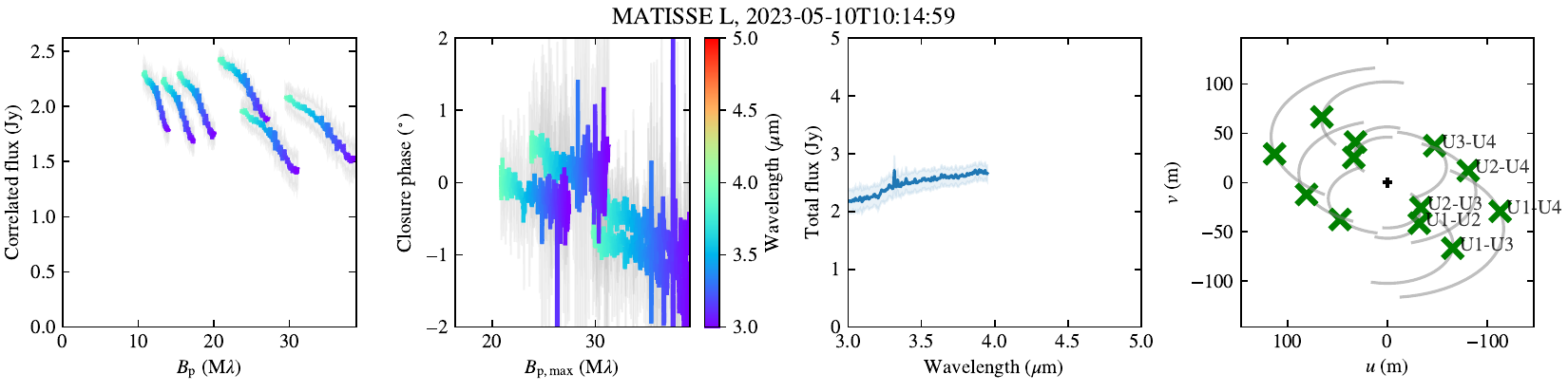}
\includegraphics[width=\hsize]{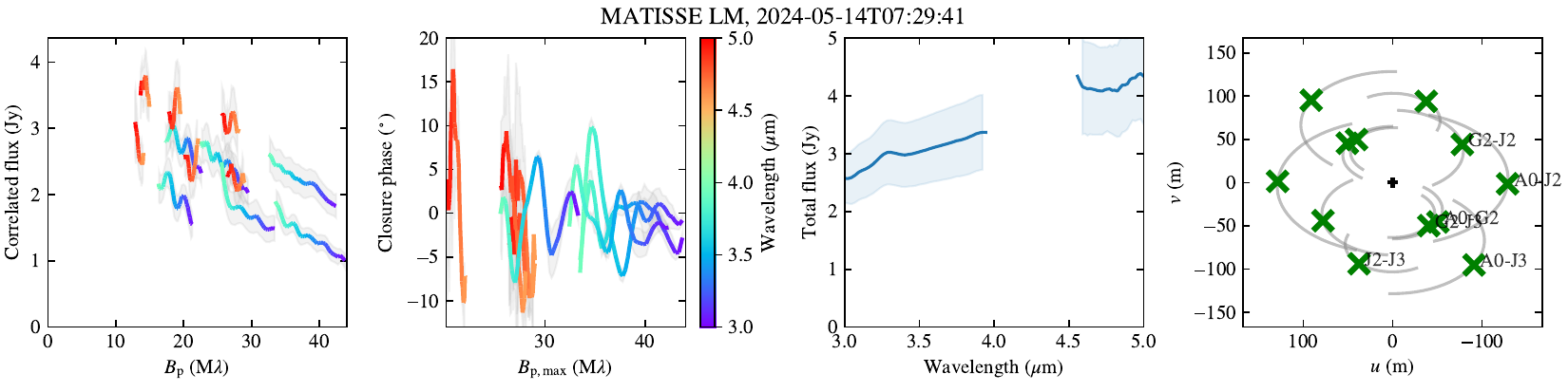}
\includegraphics[width=\hsize]{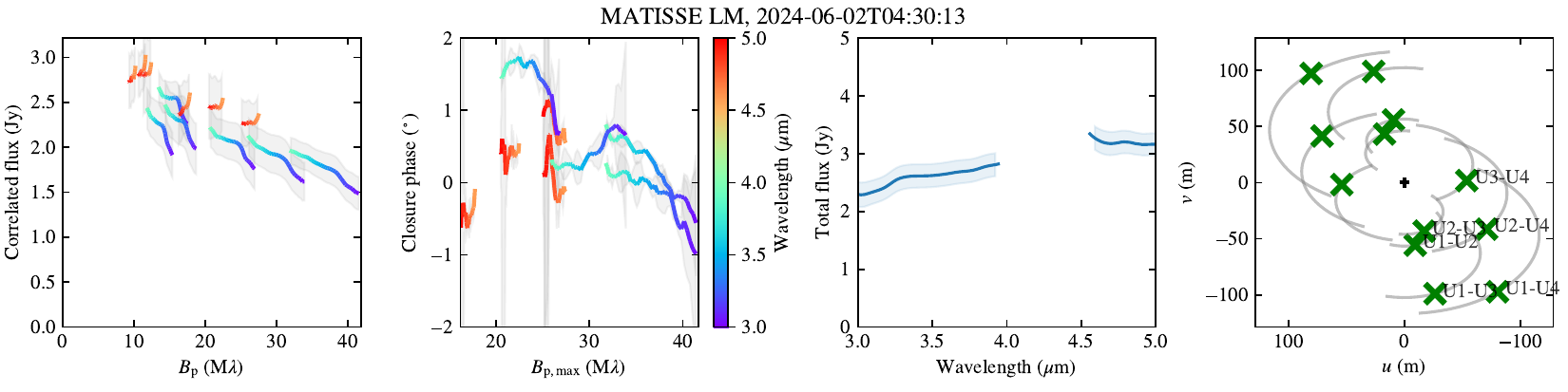}
\caption{Calibrated MATISSE $L$- and $LM$-band data sets with pipeline produced uncertainties. First column: Correlated flux as a function of the spatial frequency. Second column: Closure phase as a function of the spatial frequency corresponding to the longest baselines of the triangles. Third column: Single-dish flux as a function of the wavelength. Fourth column: $uv$ coverage of the observations. The correlated flux and the closure phase values are the average over the four non-chopped exposures of each observation, while the single-dish flux values are the average over the eight chopped exposures. The 2023 May 10 and 2024 June 2 data are UT observations, the rest are AT observations. }
\label{fig:data_L}
\end{minipage}  	 
\end{figure}

\setcounter{figure}{0}


\begin{figure*}
   \begin{minipage}{0.96\textwidth}  
    	\centering
\includegraphics[width=\hsize]{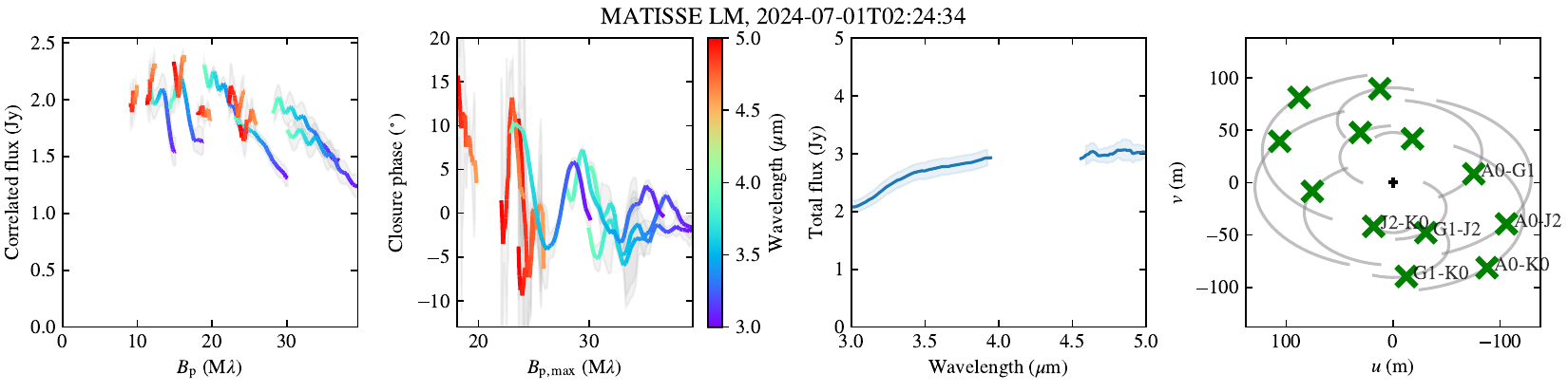}
\includegraphics[width=\hsize]{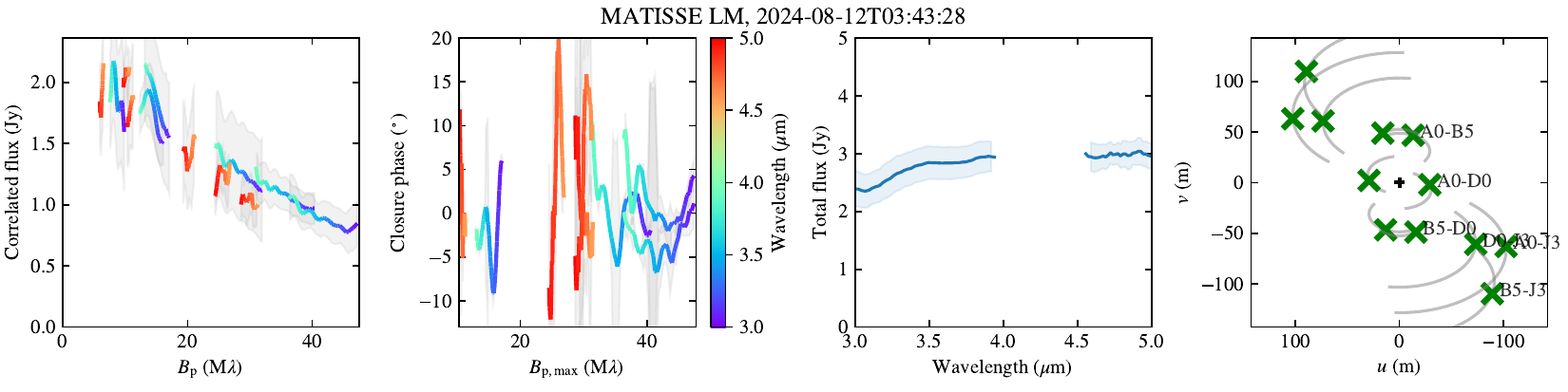}
\caption{continued.}
     	\label{fig:data_L2}
   \end{minipage}   	 
\end{figure*}

\begin{figure*}
\centering
\includegraphics[width=0.48\hsize]{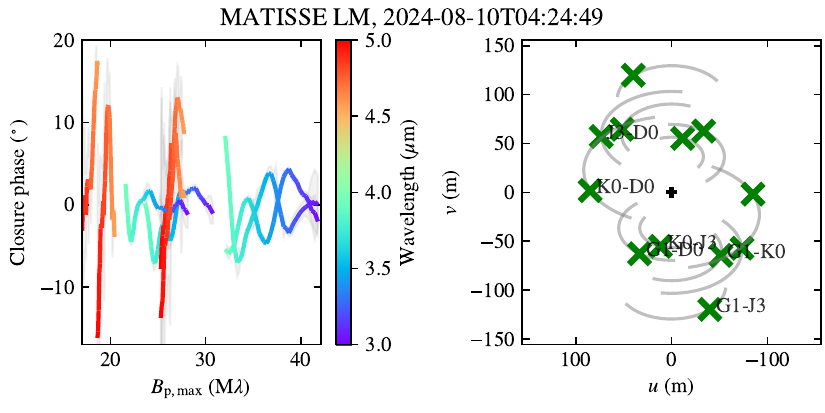}
\caption{Left panel: Average non-chopped closure phase as a function of the spatial frequency of the uncalibrated $LM$-band data set. Right panel: $uv$ coverage of the observation.}
\label{fig:data_L_uncal}
\end{figure*}

\begin{figure*}
\center
\begin{minipage}{0.96\textwidth}
\includegraphics[width=0.965\hsize]{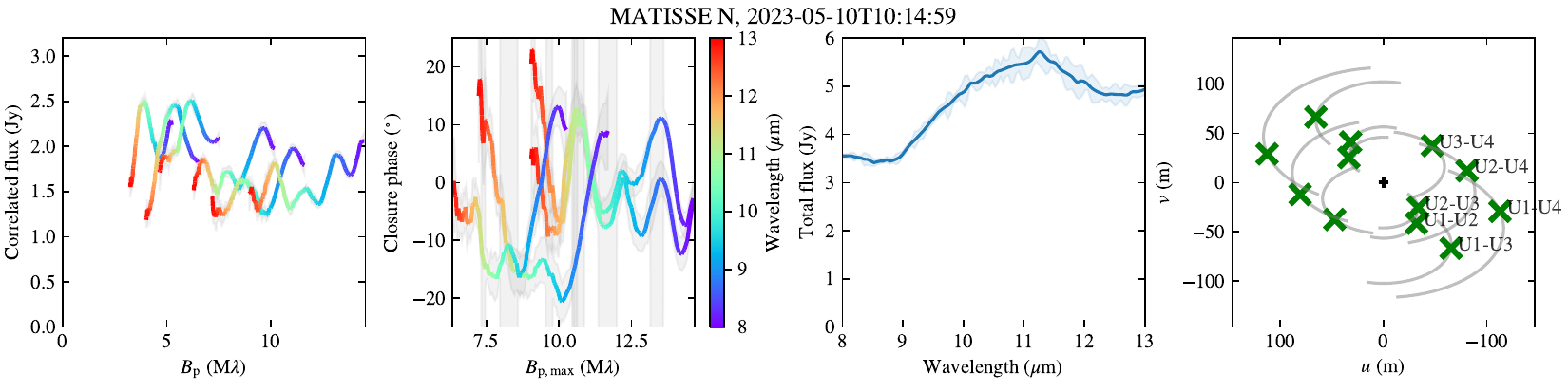}\\
\includegraphics[width=0.95\hsize]{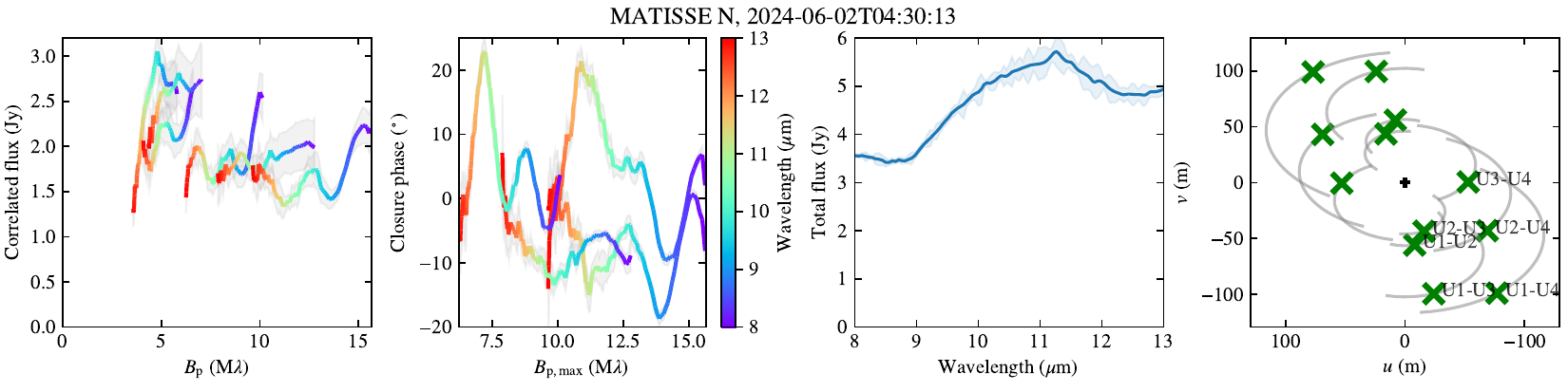}

\caption{Same as Fig.~\ref{fig:data_L}, but for the UT $N$-band data. In the third column, the single-dish flux comes from a Spitzer observation.}
     	\label{fig:data_N}
\end{minipage}	 
\end{figure*}

\FloatBarrier

\section{Additional details on the modeling}
\label{sec:app_model}
\begin{enumerate}
    \item In the first-stage fit, we use the correlated fluxes and closure phases from the average of the non-chopped exposures, and the single-dish flux from the average of the chopped exposures (as shown in Fig.~\ref{fig:data_L}). The averaging is done separately for each epoch and baseline. We fit all model parameters (FWHM, $\cos i$, and $\theta$ of the Gaussians, coordinates of the companion, and the flux contributions) except $\theta$ of the central disk. For that we use a fixed $6\degr$, that is the average of the values reported in \citet{Cugno2023} and \citet{Rigliaco2023}. We also performed test runs with $\theta$ of the central source left free, but that did not significantly improve the fit quality. The correction factors are applied on a per-epoch basis, and the resulting best-fit parameter values are taken as the ones with the lowest $\chi^2_\mathrm{red}$.

    \item In the second-stage fits, we fit the correlated fluxes and closure phases of each non-chopped exposure independently, along with the average chopped single-dish flux (per epoch). Thus, we get four set of fitted parameters per epoch. In these fits we fix the $\cos i$ and $\theta$ parameters to their values found in the first-stage modeling. The correction factors are applied on a per-baseline basis. 
\end{enumerate}

\section{Fits to the data}
\label{sec:app_fit}

\begin{table}[H]
	\caption[]{\label{tab:fit_LN_res}More detailed fit results on the FWHMs and the flux contributions of the components.}
	\small
	\centering
   	\begin{tabular}{llllll}
   \hline \hline
   Wavelength & \SI{3.1}{\um} & \SI{3.9}{\um} & \SI{8}{\um} & \SI{11}{\um} & \SI{13}{\um}\\
   \hline
   \multicolumn{6}{c}{\it Central source } \\
   \hline
   FWHM (mas) & $2.1^{+0.5}_{-0.7}$ & $2.1^{+0.5}_{-0.7}$ & $8.5^{+1.0}_{-0.8}$ & $12.8^{+1.1}_{-0.9}$ & $11.3^{+1.0}_{-0.8}$ \\[1mm]
   $F_\nu$ (Jy) & $1.7^{+0.2}_{-0.3}$ & $2.0^{+0.3}_{-0.4}$ & $3.1^{+0.3}_{-0.1}$ & $3.1^{+0.4}_{-0.2}$ & $2.1^{+0.5}_{-0.3}$ \\[1mm]
\hline
\multicolumn{6}{c}{\it Companion} \\
\hline
FWHM (mas) & $6.2^{+0.5}_{-0.8}$ & $6.2^{+0.5}_{-0.8}$ & $16.5^{+4.5}_{-1.0}$ & $19.0^{+2.6}_{-1.8}$ & $5.2^{+6.5}_{-5.0}$ \\[1mm]
$F_\nu$ (Jy) & $0.17^{+0.06}_{-0.05}$ & $0.27^{+0.10}_{-0.07}$ & $0.7^{+0.1}_{-0.2}$ & $0.6^{+0.1}_{-0.1}$ & $0.21^{+0.17}_{-0.10}$ \\[1mm]
\hline
\multicolumn{6}{c}{\it Background} \\
\hline
$F_\nu$ (Jy) & $0.5^{+0.4}_{-0.4}$ & $0.6^{+0.5}_{-0.5}$ & $0.04^{+0.00}_{-0.04}$ & $1.8^{+0.4}_{-0.4}$ & $2.5^{+0.4}_{-0.6}$ \\[1mm]
\hline
   \end{tabular}
\end{table}

\begin{figure}[t]
    	\center
\includegraphics[width=\hsize,trim={0cm 0.3cm 0cm 0.2cm},clip]{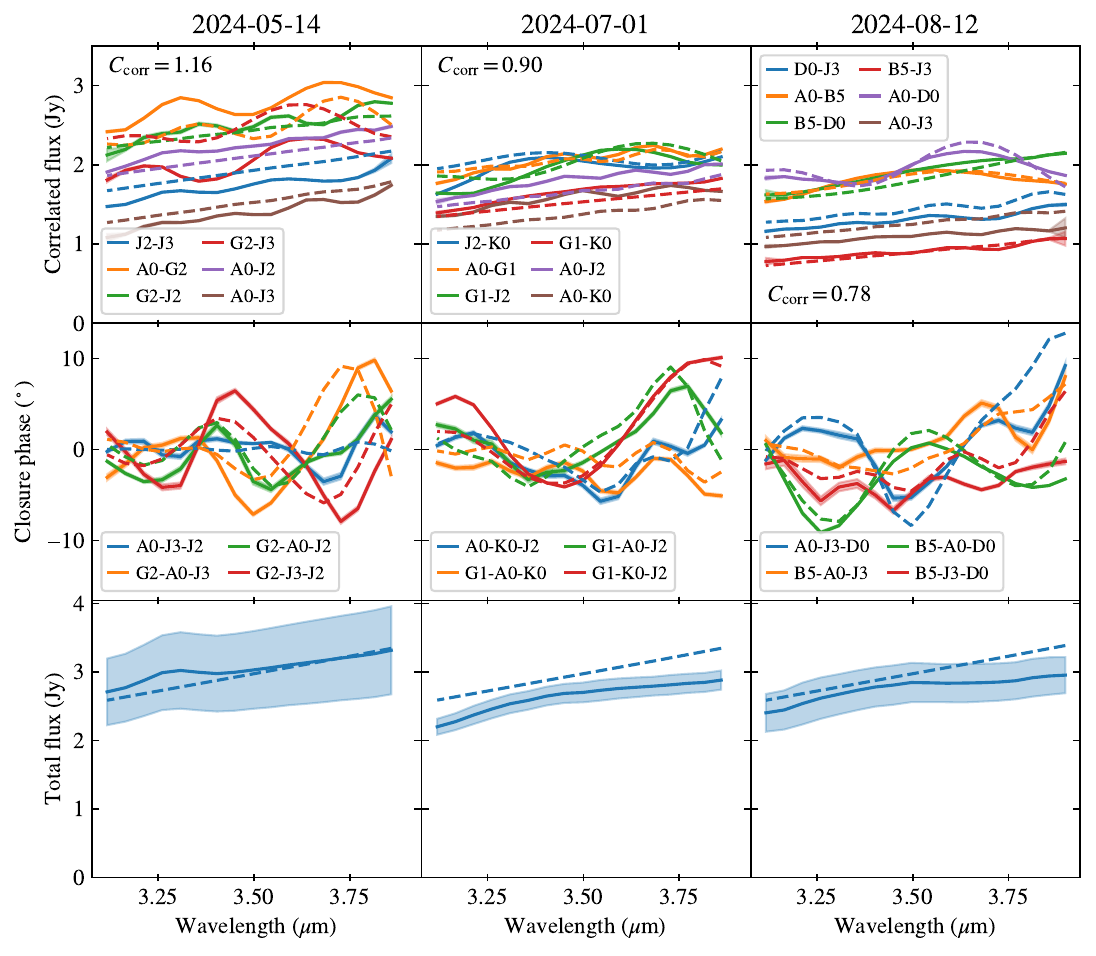}
\caption{Fits to three complete AT $L$-band data sets. Each column corresponds to a different epoch. The data are shown as solid lines, and the model as dashed lines. The correction factors for the correlated fluxes are indicated in the first row.}
     	\label{fig:fit_3epochs}   	 
\end{figure}

\begin{figure*}
\includegraphics[width=0.495\hsize]{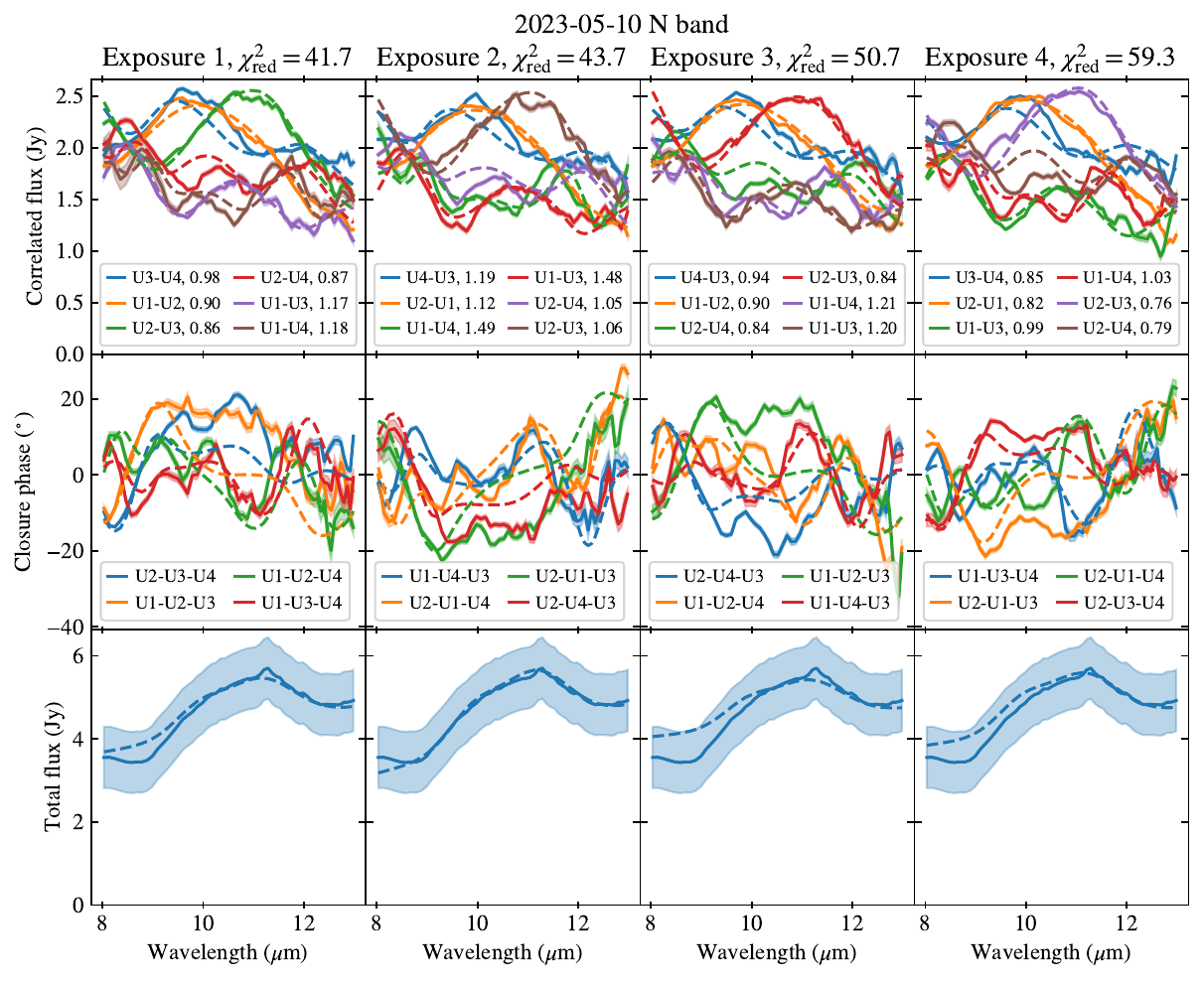}
\includegraphics[width=0.495\hsize]{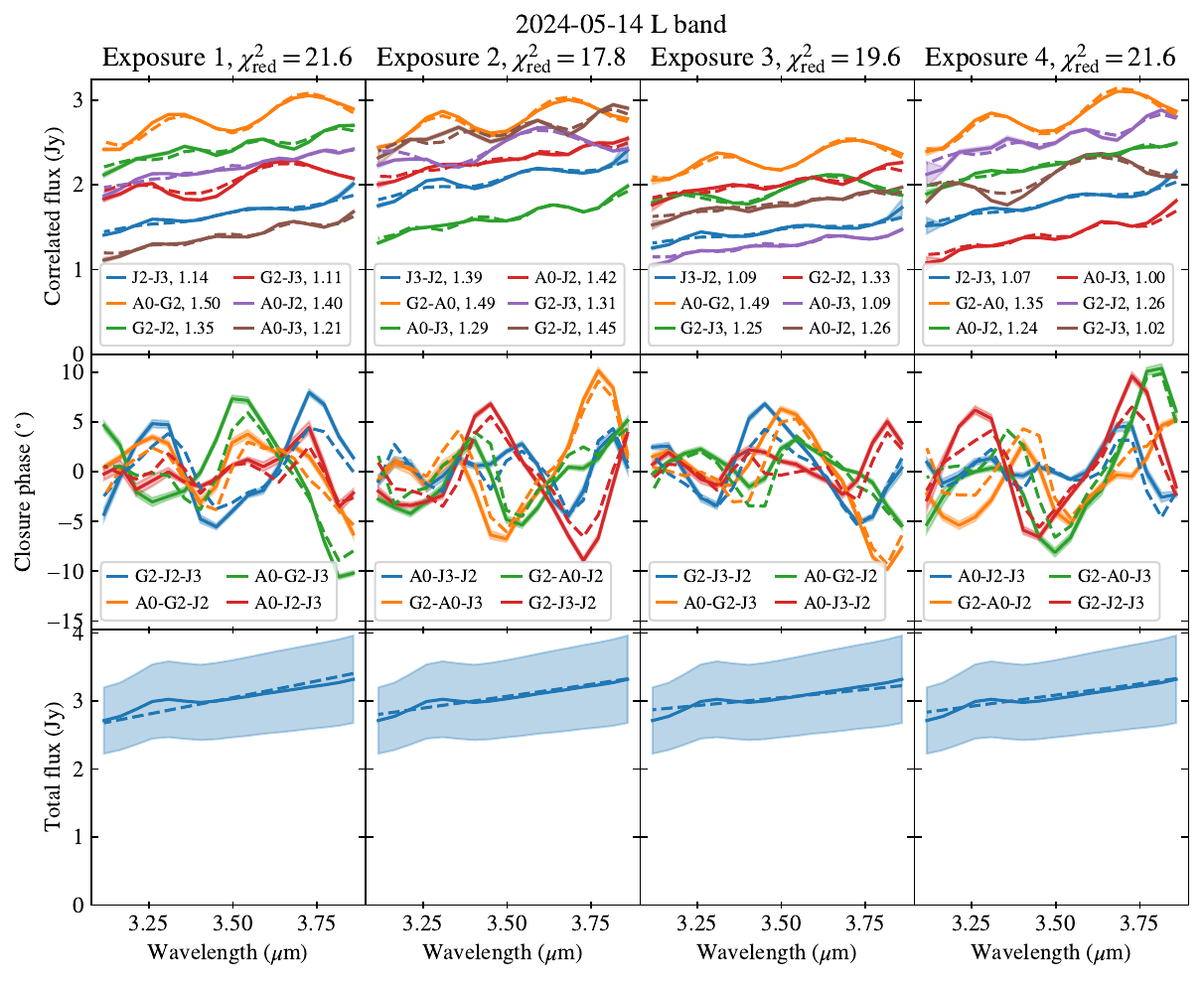}

\includegraphics[width=0.495\hsize]{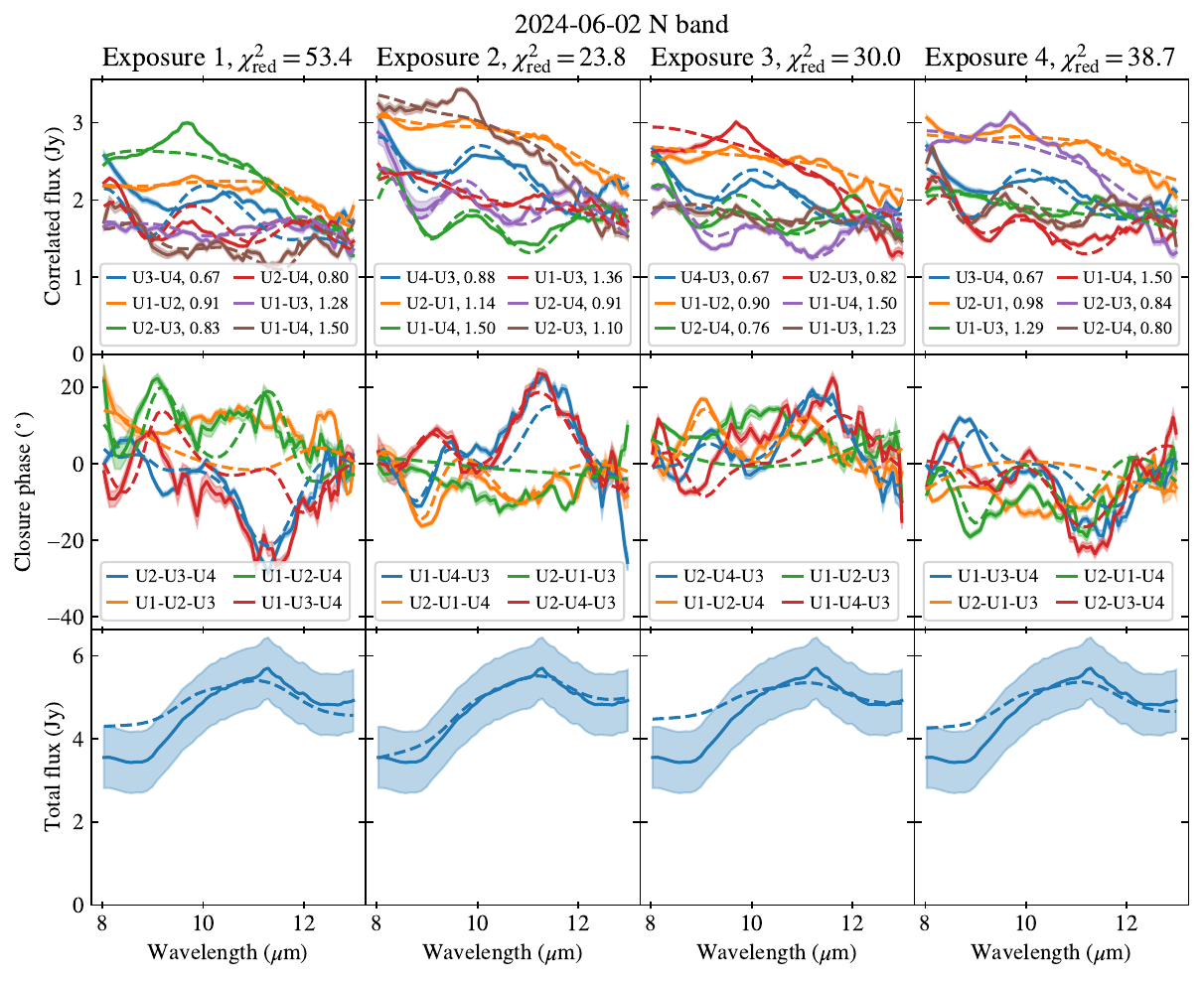}
\includegraphics[width=0.495\hsize]{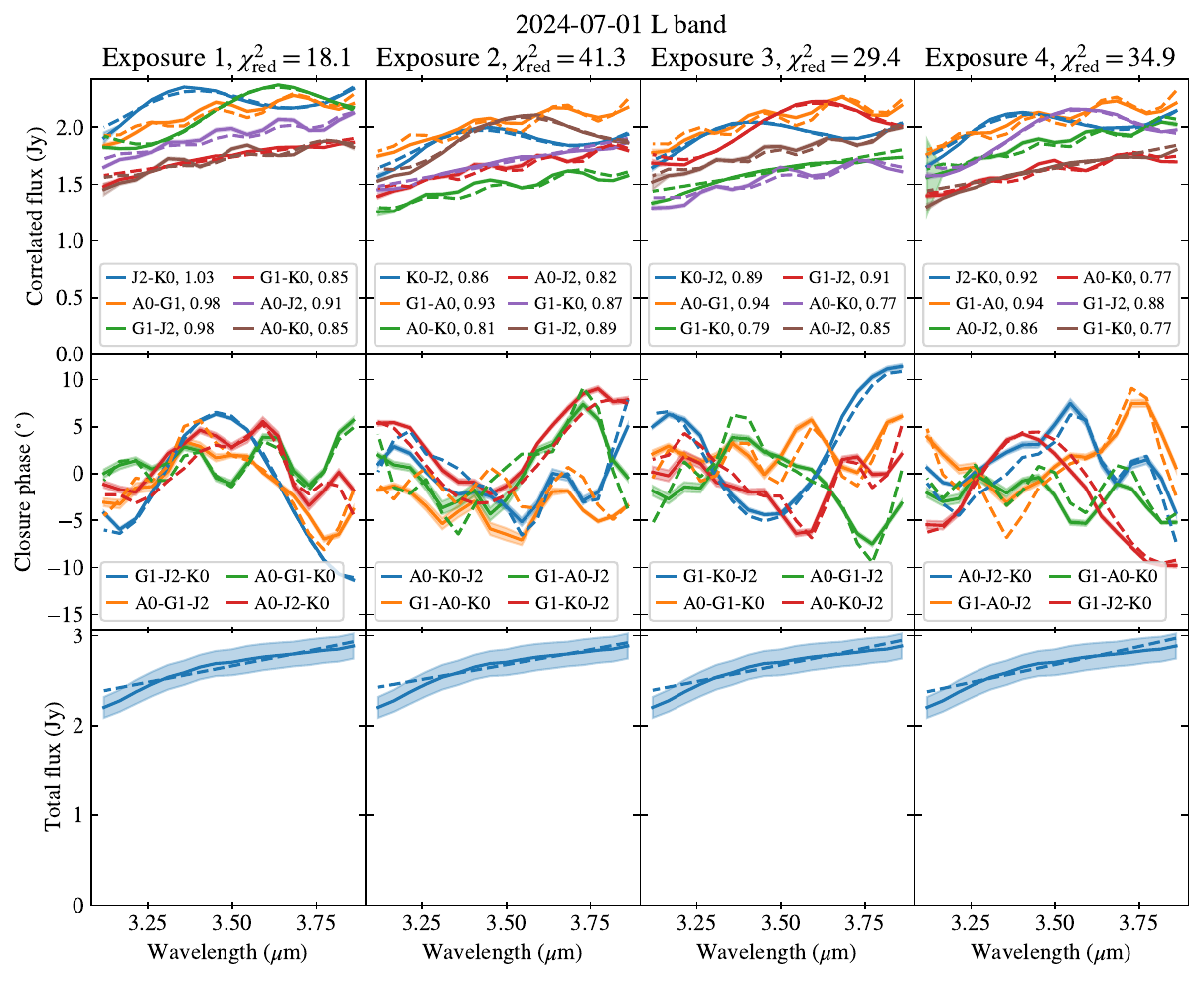}

\includegraphics[width=0.495\hsize,trim={0cm -4.7cm 0cm 0cm},clip]{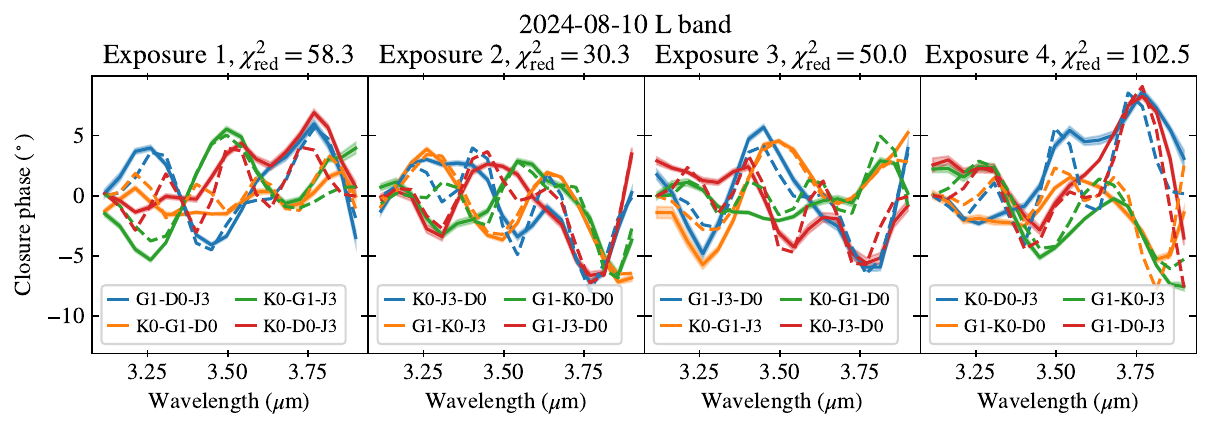}
\includegraphics[width=0.495\hsize]{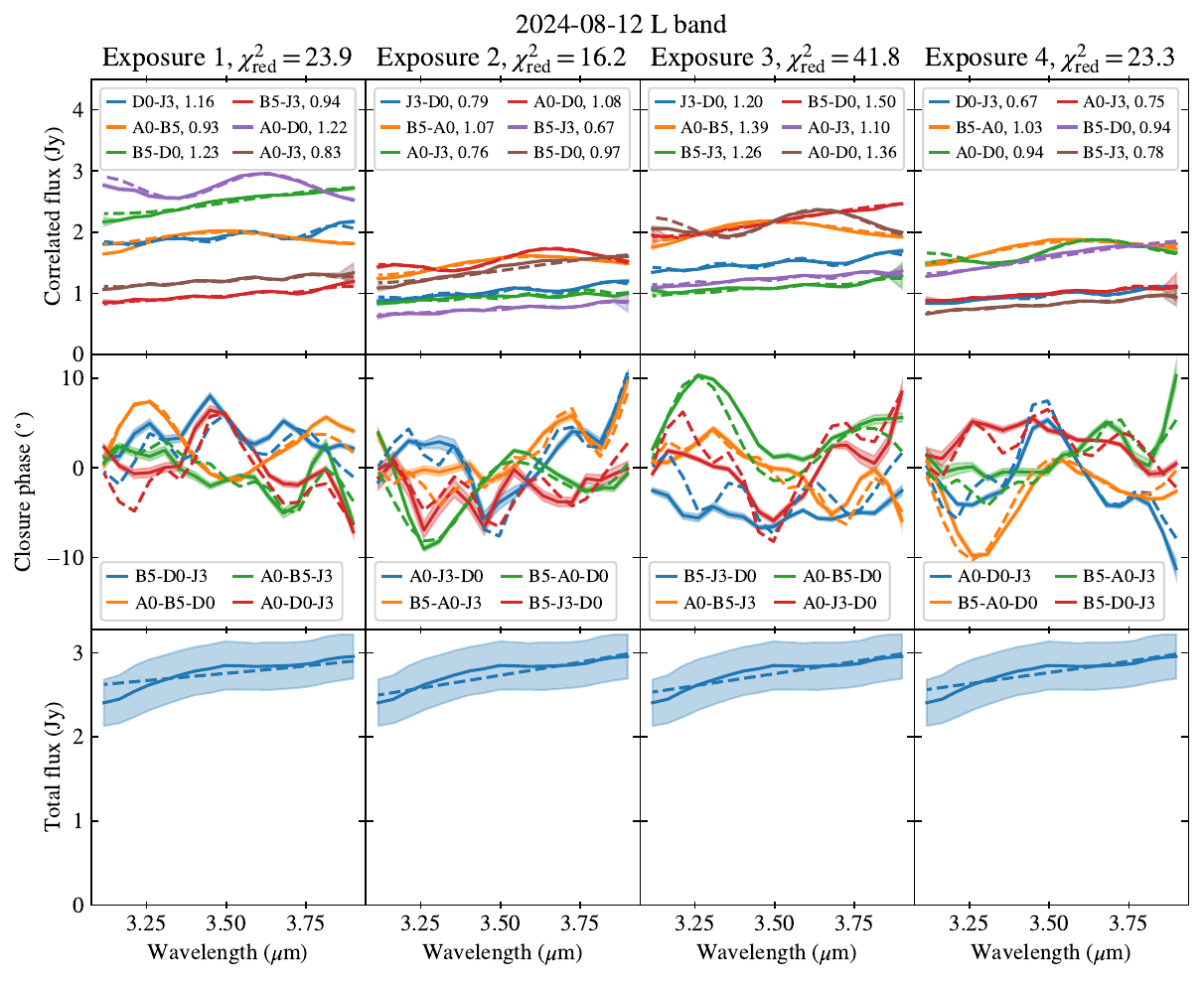}

\caption{Fits to the individual exposures in the second stage of our modeling. Each large three by four panel corresponds to a single epoch observation. A column in a panel shows the fits to a particular exposure, with the correlated flux on the top, closure phase in the middle, and single-dish flux at the bottom. The data are shown as solid lines, and the model as dashed lines. Correction factors are indicated in the legend of the correlated flux plots, for each baseline. The 2024 Aug 10 data set is incomplete, and thus only the closure phases were fitted.}
     	\label{fig:fit_CF_CP}
\end{figure*}

\end{appendix}

\end{document}